\documentclass[
 reprint,
superscriptaddress,
 amsmath,amssymb,
 aps,
pra,
longbibliography
]{revtex4-2}

\usepackage{graphicx}
\usepackage{dcolumn}
\usepackage{bm}
\usepackage{hyperref}
\usepackage{notes2bib}
\usepackage[nice]{nicefrac}
\usepackage[normalem]{ulem}
\usepackage{xcolor}
\usepackage{physics}
\usepackage{mathtools}
\newcommand{\um}[1]{\,\mathrm{#1}}

\usepackage{amssymb}
\usepackage[a4paper, total={6.5in, 9in}]{geometry}

\usepackage{graphicx}
\usepackage{amsthm}
\usepackage{amsmath}
\newcommand{\e}{\mathrm{e}}
\newcommand{\ii}{\mathrm{i}}

\newcommand{\sinc}{\mathrm{sinc}}

\begin{document}

\title{Momentum-resolved two photon interference of weak coherent states}

\author{Fabrizio Sgobba}
\affiliation{Dipartimento Interateneo di Fisica, Universit\`a di Bari, 70126 Bari, Italy}
\affiliation{Istituto Nazionale di Fisica Nucleare (INFN), Sezione di Bari, 70126 Bari, Italy}

\author{Francesco Di Lena}
\affiliation{Agenzia Spaziale Italiana - Matera Space Center, Contrada Terlecchia snc, 75100 Matera, Italy}

\author{Danilo Triggiani}
\affiliation{Istituto Nazionale di Fisica Nucleare (INFN), Sezione di Bari, 70126 Bari, Italy}
\affiliation{Dipartimento Interateneo di Fisica, Politecnico di Bari, 70126 Bari, Italy}

\author{Deborah Katia Pallotti}
\affiliation{Agenzia Spaziale Italiana - Matera Space Center, Contrada Terlecchia snc, 75100 Matera, Italy}

\author{Cosmo Lupo}
\affiliation{Dipartimento Interateneo di Fisica, Universit\`a di Bari, 70126 Bari, Italy}
\affiliation{Istituto Nazionale di Fisica Nucleare (INFN), Sezione di Bari, 70126 Bari, Italy}
\affiliation{Dipartimento Interateneo di Fisica, Politecnico di Bari, 70126 Bari, Italy}

\author{Piergiorgio Daniele}
\affiliation{Dipartimento di Elettronica, Informazione e Bioingegneria, Politecnico di Milano, 
Piazza Leonardo da Vinci 32, 20133 Milano, Italy}

\author{Gennaro Fratta}
\affiliation{Dipartimento di Elettronica, Informazione e Bioingegneria, Politecnico di Milano, 
Piazza Leonardo da Vinci 32, 20133 Milano, Italy}

\author{Giulia Acconcia}
\affiliation{Dipartimento di Elettronica, Informazione e Bioingegneria, Politecnico di Milano, 
Piazza Leonardo da Vinci 32, 20133 Milano, Italy}

\author{Ivan Rech}
\affiliation{Dipartimento di Elettronica, Informazione e Bioingegneria, Politecnico di Milano, 
Piazza Leonardo da Vinci 32, 20133 Milano, Italy}

\author{Luigi Santamaria Amato}
\email{luigi.santamaria@asi.it}
\affiliation{Agenzia Spaziale Italiana - Matera Space Center, Contrada Terlecchia snc, 75100 Matera, Italy}

\keywords{Quantum Information, Interferometry, Quantum sensing}

\begin{abstract}
We demonstrate an experimental scheme for high-precision position measurements based on transverse-momentum-resolved two-photon interferometry with independent photons and single photon avalanche diode  (SPAD) arrays. 
Our scheme extends the operative range of Hong-Ou-Mandel interferometry beyond its intrinsic constraints due to photons indistinguishability, paving the way to applications in high-resolution imaging. 
We assess the experimental results against the ultimate precision bounds as determined by quantum estimation theory.
Our experiment ultimately proves that transverse-momentum resolved measurements of fourth-order correlations in the fields can be employed to overcome spatial distinguishability between independent photons. The relevance of our results extends beyond sensing and imaging towards quantum information processing, as we show that partial photon distinguishability and entanglement impurity are not necessarily a nuisance in a technique that relies on two-photon interference.
\end{abstract}

\maketitle
\onecolumngrid 

\section{Introduction}
\label{sec:intro}

Dirac once claimed that ``interference between different photons never occurs''~\cite{dirac1958principles}. 
The statement was soon challenged by Mandel and Magyar~\cite{magyar1963}, who experimentally demonstrated that photons can, in fact, interfere with one another.
Contrary to self-interference, multi-photon interference is a purely quantum phenomenon that has no classical counterpart.
As a consequence, in the context of quantum information science, it gives access to information that is precluded to any classical approach.
Multi-photon interference is best appreciated in Hong-Ou-Mandel (HOM) interferometry~\cite{Hong1987}. 
HOM interference consists in a drop in coincidence events measured between detectors placed at the output ports of a balanced beam splitter, given that two indistinguishable photons enter each from one of the two input ports. 
The probability amplitudes of a combined detection interfere destructively, producing the said drop in coincidence events as a result of the \emph{which-path} uncertainty, with visibility crucially depending on photon indistinguishability.

The majority of experiments based on HOM interference are performed using the frequency-correlated two-photon state (FCTS) as the probe. This state ensures that a single photon enters each input port of the beam splitter, allowing for a maximum visibility of 100\% in the case of complete indistinguishability.  The HOM experiment is very general and can be implemented using various probe states, such two-photon squeezed state (TPQS), coherent states or thermal states. 
A TPQS state can be generated by the same nonlinear processes (e.g.: spontaneous parametric downconversion) used for FCTS but operated in the higher-gain regime. This state includes higher-order photon-pair terms ($|2,2\rangle$, $|3,3\rangle$, etc.), which reduce the interference visibility due to multiphoton contributions. 
When a weak coherent state is used as the probe state ~\cite{Chen2016}, multiphoton contributions and Poissonian photon-number statistics reduce the maximum achievable visibility to 50\% in the case of perfect distinguishability. In general, two weak coherent states coming from the same source will have a fixed phase relation between photons, resulting in a first order interference that ultimately masks the second order interference that one aims to investigate.  Becomes therefore mandatory, while working with coherent state probes, to suppress first order interference by randomizing the phase. Finally if thermal states are adopted~\cite{Ihn2017}, the maximum obtainable visibility is 33\%.

In a typycal HOM interferometer by
introducing a variable degree of distinguishability, as for example a mismatch $\Delta t$ in the arrival time, it is possible to observe a coincidence profile. If the coincidences events $C(\Delta t)$ are measured at different values of $\Delta t$, it is found that $C(\Delta t) \to 0$ for $\Delta t/\tau \to 0$ 
(HOM dip),
whereas $C(\Delta t) \to C_0$ for $\Delta t/\tau \gg 1$. 
Here $C_0$ is the background value of coincidence events registered for non-interfering photons, and $\tau$ is the biphoton temporal coherence. 
HOM interference can be exploited in delay measurements to achieve detection limits well below the time jitter of the detectors employed, even below the order of magnitude of the temporal coherence $\tau$ itself~\cite{dauler1999}.
Advances in light engineering~\cite{Walborn2003,DAmbrosio2019, Kim2017} and single-photon detection systems~\cite{DelloRusso2022} have sparked a new wave of HOM interference-based experiments~\cite{kim2020hong, Kobayashi2016,kim2020two}, with application in e.g.~quantum information science~\cite{stobinska2019quantum}, quantum imaging~\cite{kolenderska2020fourier, hayama2022high},
quantum metrology~\cite{Lyons2018, Harnchaiwat2020, brady2021frame,Sgobba2024}.

Standard HOM interferometry operates within the constraint that $\Delta t< \tau$, which guarantees at least a partial degree of photon indistinguishability.
To overcome this limitation one resorts to conjugate-variable resolution techniques.
For example, in a measurement of time delay, one considers frequency-resolved coincidence. 
The count of coincidence events is performed here in the frequency domain, within a resolution $\delta_{\omega}$ specific to the system employed, and a quantum beat of period proportional to $\Delta t^{-1}$ is observed. 
It was shown that, in this configuration, the new operational range ($\tau_R$) is determined by the Heisenberg uncertainty principle as $\tau_R\simeq 1 /\delta_{\omega}>\tau$~\cite{Triggiani2023,Dilena2025}.
This approach has been investigated since the early years 2000's~\cite{Legero2004,Jin2015,Gerrits2015}, and later applied to quantum communications~\cite{PietxCasas2023}, quantum coherence tomography~\cite{IbarraBorja2019,YepizGraciano2020}, quantum enhanced imaging~\cite{Parniak2018, Muratore2025}, entangled pair production~\cite{Zhao2014,Jin2016}, boson sampling~\cite{Orre2019,Tamma2015,Wang2018}, coalescence states~\cite{schiano2024}, and in precision metrology~\cite{Parniak2018,Chen2023,Triggiani2024,Triggiani2025,Maggio2025}.  

In this work we experimentally demonstrate the resolution technique for the position-momentum conjugate variables, hence implementing a recent theoretical proposal by Triggiani and Tamma~\cite{Triggiani2024}~\footnote{While finalizing the manuscript, similar results were independently reported in~\cite{wv2d-5z4g}.
	Ref.~\cite{wv2d-5z4g} employs entangled SPDC Type-II photons at telecom wavelength and transverse-momentum filtering.
	By contrast, here we use independent photons at optical frequency from coherent light, which paves the way to imaging applications.
	Furthermore, our scheme achieve momentum resolution using a SPAD arrays instead of filtering, hence yielding a much more time-saving and information-efficient way to collect data.
}.
In our setup, we aim at a precision measurement of a position displacement $\Delta x$ through the observation of the coincidence counts $C_{k}(\Delta x)$, given a value $k$ of the transverse momentum. 
By resolving the coincidence events in momentum space we are able to retrieve the interference phenomenon even when the two wave packets are completely distinguishable due to their high transverse separation, i.e.~$\Delta x > \sigma_x$, where
$\sigma_x$ is the biphoton waist, thus extending the operative range of the HOM interferometer.

Here we demonstrate a reliable technique for conjugate-variable passive resolution.
Our approach will find applications in high-resolution imaging, as it overcomes some of the limitations of direct imaging due to finite pixel pitch.
In fact, a tiny change in displacement, which causes a proportionally small translation in the intensity distribution with direct imaging, causes in our approach a more easily visible change in the beating oscillation in the momentum space.
More generally, our work demonstrates that it is possible to observe quantum beats, and therefore two-photon quantum interference, between independent photons whose wave packets do not overlap.
Other than applications in biosensing and imaging, such as correlation plenoptic imaging~\cite{d2016correlation,di2020correlation,pepe2017diffraction}, it is possible to envision an extension of this technique to quantum communication and to quantum information processing with partially distinguishable photons.

\section{Model}
\label{sec:model}
Consider two independent photons impinging on the two faces of a balanced beam splitter.
For example, we can think the two photons as emitted by two single-photon sources (or two weak coherent states, as in our setup), where one labels the positions of an object on a plane, such as a molecule in a biological sample, and the other is used as reference.
If single-photon cameras are employed in the far field at the output ports of the beam splitter, and bunching and antibunching events are registered for each pair of pixels, quantum beats with frequency of oscillation proportional to the transverse separation $\Delta x$ between the two sources can be observed~\cite{Triggiani2024}.
In particular, assuming a sufficiently high resolution of the cameras, quantified by the transverse-momentum sensitivity $\delta$, so that  $\delta\ll1/\Delta x$ and $\delta\ll1/2\sigma_x$, where  $1/2\sigma_x \equiv \sigma_k$ is the transverse momentum photon wave packet width at the detection, the probability of observing the two photons antibunching (A) or bunching (B) having transverse momenta $k_1,k_2$ reads
\begin{equation}
	P_{A/B}(k_{1},k_{2})= \frac{1}{2}f(k_{1})f(k_{2})\left(1\mp\mathcal{V}\cos((k_{1}-k_{2})\Delta x)\right) \, ,
	\label{eq:InfRes}
\end{equation}
where $f$ is the transverse-momentum distribution of the single photon~\cite{Triggiani2024} and $\mathcal{V}$  represents the interference visibility.
To take into consideration a finite minimum detectable change in transverse momentum $\delta$, caused for example by the finite width of the pixels, we need to integrate Eq.~\eqref{eq:InfRes} over finite intervals of size $\delta$, yielding a discrete probability function $P_{A/B}^{k_{01},k_{02}}=\int_{k_{01}-\frac{\delta}{2}}^{k_{01}+\frac{\delta}{2}}\int_{k_{02}-\frac{\delta}{2}}^{k_{02}+\frac{\delta}{2}}\dd k_1\dd k_2\ P_{A/B}(k_{1},k_{2})$, where $k_{01},k_{02}$ are the detectable transverse momenta.
By measuring with a CCD and an intense laser  $\sigma_x\simeq 35 \mu m$,  we can assume that within such intervals $f(k)$ remains approximately constant, i.e. for $\delta*\sigma_x\ll 1$
the integration of the probability in Eq.~\eqref{eq:InfRes} simplifies to
\begin{equation}
	P_{A/B}^{k_{01},k_{02}}\simeq \frac{1}{2}f(k_{01})f(k_{02})
    \int_{k_{01}-\frac{\delta}{2}}^{k_{01}+\frac{\delta}{2}}\int_{k_{02}-\frac{\delta}{2}}^{k_{02}+\frac{\delta}{2}}\dd k_1\dd k_2\ (1\mp\mathcal{V}\cos((k_{1}-k_{2})\Delta x) \, ,
    \label{eq:IntermediateProb}
\end{equation}
that can be easily evaluated as
\begin{equation}
	P_{A/B}^{k_{01},k_{02}}\simeq \frac{C}{2}\left(1\mp\mathcal{V}\,\sinc^2\left(\frac{\Delta x \delta}{2}\right)\cos((k_{01}-k_{02})\Delta x)\right),
    \label{eq:ProbTransv}
\end{equation}
with $C=f(k_{01})f(k_{02})\delta^2$, while the $\sinc$ function intuitively arises as the Fourier transform of the rectangular distribution associated with the integration domain of each pixel.

We see from Eq.~\eqref{eq:ProbTransv} that a small variation of $\Delta x$ does not translate the probability distribution observed in the $k$-domain, instead it varies the beating oscillation in $k_{01}-k_{02}$.

From Eq.~\eqref{eq:ProbTransv} it is possible to assess the sensitivity of this measurement scheme for the estimation of the displacement $\Delta x$, and compare it with the ultimate precision achievable in nature.
This can be done by evaluating the Fisher information~\cite{Cramer1999}, denoted $F$, and the Cramér-Rao bound associated with the transverse-momentum resolved measurement, and compare it with the quantum Fisher information~\cite{Helstrom1969,Liu2020}, denoted $H$.
The quantum Cramér-Rao bound after $N$ repeated measurements is
\begin{equation}
	\delta\Delta x_{Q}=\frac{1}{\sqrt{N H}}=\sqrt{\frac{2}{N}} \, \sigma_x \, ,
	\label{eq:QFI}
\end{equation}
since $H=1/(2\sigma_x^2)$ (see appendix \ref{sec:appendix}), where $\sigma_x$ is the transversal width of the wave-packet. Eq.~\eqref{eq:QFI} implies the feature of photons with narrower transversal wave-packets to yield a greater sensitivity for the estimation of the displacement $\Delta x$, and, remarkably, such a sensitivity does not depend on the transverse separation to be estimated.  $H=1/(2\sigma_x^2)$, was calculated in ~\cite{Triggiani2024} for single photons, here in the appendix, it is shown that its value remains unchanged when coherent states are used as probe.
The highest precision achievable with a given measurement scheme is instead given by the classical Cramér-Rao bound~\cite{Cramer1999},
\begin{equation}
	\delta\Delta x=\frac{1}{\sqrt{N F}} \, ,
\end{equation}
where $F\leqslant H$ is the Fisher information associated with the measurement. The Fisher information can be calculated from the probability distribution $P(x;\Delta x)$ associated with the measurement outcome $x$ as $F = \mathbb{E}_X\left[\left(\frac{\partial}{\partial\phi}\ln P(X|\phi)\right)^2\right]$, where $\mathbb{E}_X$ denotes the expectation value over the probability distribution $P$ (see appendix \ref{app:derivation}).

It is known that transverse-momentum resolved HOM interference, achievable though cameras positioned in the far-field regime can saturate such a precision when the photons are indistinguishable in any non-spatial degrees of freedom so that the visibility of the interference is $\mathcal{V}=1$, and the resolution of the cameras, quantified by the transverse-momentum sensitivity $\delta$, is sufficiently high so that $\delta\ll1/2\sigma_x$ and $\delta\ll1/\Delta x$~\cite{Triggiani2024}.
For partially distinguishable photons $\mathcal{V}<1$ it is still possible to estimate separation for arbitrarily separated photons with the spatially resolved detection scheme.

\begin{figure}
    \centering
	\includegraphics[width=0.55\linewidth]{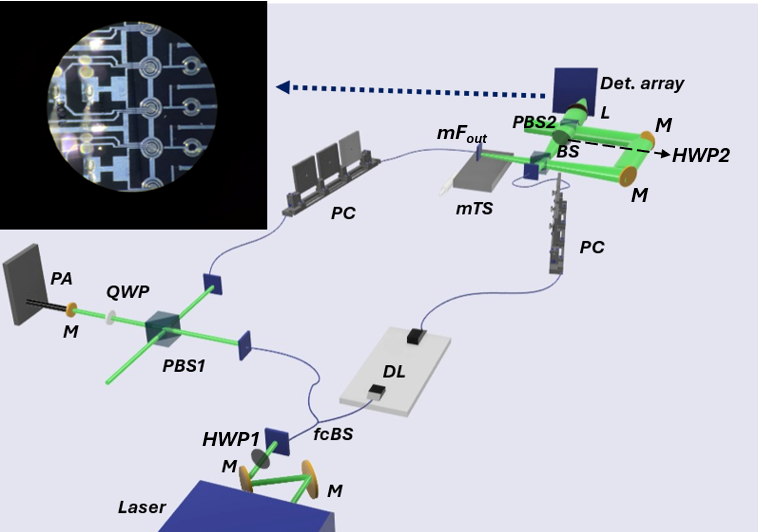}
	\caption{\label{fig:stp}3D representation of the optical setup devised and employed for the experiment. {M}= Mirrors, {PBS} =polarizing beam splitter ({fc}= fiber-coupled) and {BS} = non-polarizing beam splitter, {PA} = piezo actuator,  {HWP1 and HWP2} = half-wave plates, {QWP} = quarter-wave plate, {DL}= optical delay line, {PC}= polarization controllers, {mTS}= motorized translation stage,  {L}= optical lens and $mF_{out}$ is the collimator placed on the translation stage used to shift transverse position of one of the impinging photons. In the inset, image of the single photon avalanche diodes (SPADs) array acquired via a microscope.
	}
\end{figure}

\section{Methods and Apparatus}
\label{sec:ma}
The experimental apparatus consists of a free-space Mach-Zehnder interferometer (MZI) illuminated by a pulsed weak coherent source. A collimator mounted on a transverse translation stage ensures control over the spatial overlap between interfering beams at the beam splitter. Both outputs undergo free-space diffraction before being detected in far-field conditions, making the final array of detectors momentum sensitive. The source is optically low-pass filtered in order to wash out $G^{(1)}$ interference via phase randomization (figure \ref{fig:stp}).
This is necessary since in the ordinary HOM interference one single photon (Fock state $|1\rangle$) enters each of the two BS inputs. For these states the uncertainty in the photon number is $\Delta n = 0$ while the phase is entirely random.
On the other hand, classical interference (interference of the fields) occurs when relative phase between input is determined and fixed.
So, when coherent states are used, relative phase between photons impinging on the second Mach Zehnder beam splitter (BS in figure 1 that represents the BS of HOM experiment), is determined and fixed because the path length difference between 
MZ arms is fixed. For this reason, two photon interference is masked by the fringes of classical interference. To smear out the classical interference fringes and make two photons interference appear, it is necessary to acquire the coincidence 
by averaging the relative phase between impinging photons.

The weak coherent source is constituted by a pulsed laser with a central frequency of $531.5\um{nm}$ and linewidth $0.5\um{nm}$, resulting in a pulse width of $\sim 2\um{ps}$. Its repetition rate is $40\um{MHz}$. 
After an attenuator and a half-wave plate, the laser light is coupled into a single-mode fiber and split via a fiber-coupled beam splitter (fcBS). A half-wave plate (HWP1)  before fcBS allows us to align linear polarization with PM fiber axis. 
One of the two outputs of fcBS encounters an optical delay line (DL), undergoes a variable path in free space and is later coupled back in fiber. 
The task of the DL is to optimize the temporal indistinguishability between the arrival times of the photons impinging on BS, hence ensuring the highest possible visibility for the $G^{(2)}$ interference.
The other output is transmitted through a free-space polarizing beam splitter (PBS1) and a quarter-wave plate (QWP) and then reflected back via a mirror glued to a piezoelectric actuator (PA). 
The system described allows for the low-pass optical filtering of $G^{(1)}$ interference, performed by feeding to PA a $1\um{kHz}$ symmetric triangle wave amplified to a peak-to-peak voltage of $V_\text{pp}=30\um{V}$ via an amplification stage. The piezo-induced vibration spans a  $\approx 32\cdot\lambda/2$-long optical path. 
Ideally, if the piezo were perfectly calibrated, to uniformly sample $G^1$ interference fringe during the measurement 
time would suffice to span a total length of  $\lambda/4$ per ramp, and that in turn would shift the phases in the
 range [0, 2 $\lambda/4= \lambda/2$] due to the roundtrip. If the piezo has even a slight miscalibration, however, a longer
or shorter strain with respect to $\lambda/4$ would ultimately result in a residual $G^1$ visibility. Our solution was to
 minimize any possible systematic error, by employing a piezo with a much larger strain  ($\sim$ 32 times larger than $\lambda/4$ ), 
thus spreading out and dividing any systematic miscalibration by a factor of $\sim$ 32.

The optically filtered back-reflection, passing a second time through the QWP has the correct polarization to be fully reflected by the PBS into the exit mode, where it is again coupled into an optical fiber. The polarizations of the two beams expected to be mixed in BS are made indistinguishable by polarization controllers (PC).

One of the two fiber collimators feeding into the input ports of BS is mounted on a linear motorized translation stage (mTS) designed to move in the direction orthogonal to the optical axis, reliably varying the spatial overlap of the optical paths before mixing and consequently tuning their axial distinguishability.  
The states coming out of BS, which have undergone a pathway mixing, propagate in free space for two different lengths. One of the two outputs of BS is in fact coupled back on the other mode after a $6 \um{ns}$-long detour with a combining free space polarizing beam splitter (PBS2) and a half wave plate (HWP2)   to balance the photon counts coming from each arm on the final detector. In this way both outputs of BS can be measured by the same detection array, in different times well within the repetition rate of the source ($25 \um{ns}$), acting as a clock.  A $f= 300 \um{mm}$ plano-convex lens (L) is employed to move the image plane on the detector, ensuring that the measurement takes place in far-field conditions.

The detection system is a linear array of 8 single photon avalanche diodes (SPADs) connected to a single FPGA, built within the facilities of Politecnico di Milano. The center-to-center distance between two adjacent SPADs is $250\um{\mu m}$. However, the theoretical transverse-momentum sensitivity $\delta_{th}$ can be retrieved by their width $w=50\um{\mu m}$ through the formula $\delta_{th}\simeq 2\pi w/\lambda f\simeq 2$mm$^{-1}$, with $f$ focal distance of the lens and $\lambda$ central wavelength of the photons.
The detector is synced with the $40 \um{MHz}$ TTL reference coming from the laser head. Each laser pulse acts as a trigger for a 14-bit time to amplitude conversion ramp (TAC)  $25\um{ns}$-wide. Using a reference clock with a periodicity that corresponds to the TAC duration allows us to maximize time-tagging resolution. In these conditions, the tagging resolution of each SPAD is $25\um{ns}/2^{14}\approx1.5\um{ps}$, comparable to the pulse width of the source.

\begin{figure*}
	\includegraphics[width=1\linewidth]{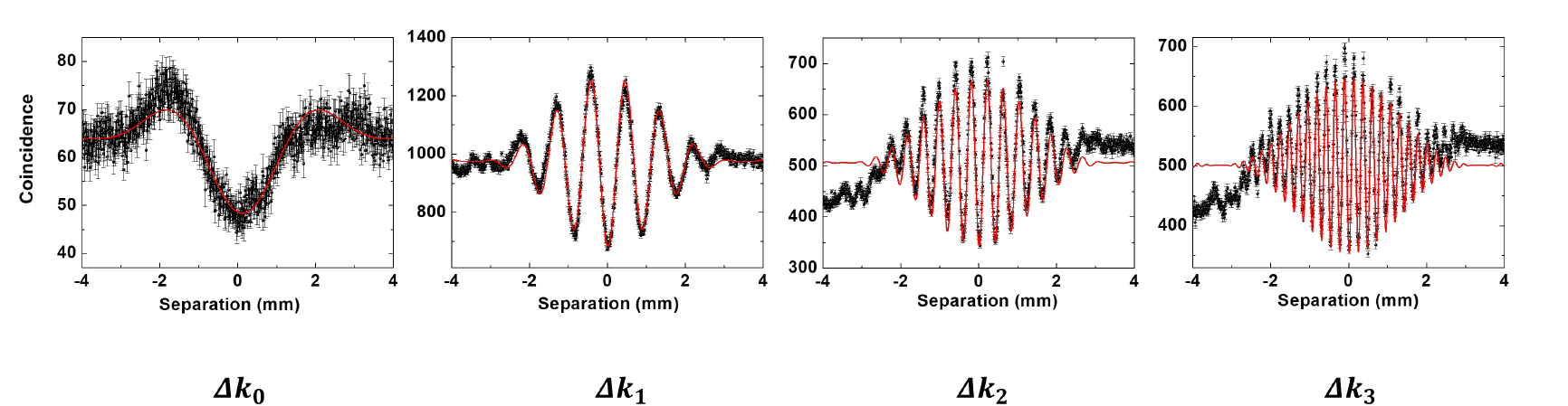}
	\caption{\label{MR HOM} \textbf{Momentum resolved anti-bunching.}
		Black points are the measured coincidences ($C^{A}_{i,j}$) for four different pixel pairs $i,j$ corresponding to the wavevector x-projections ($k_i,k_j$). The four plots show coincidence quantum beats $C^{A}_{i,j}$ between pixels corresponding to projections difference $\Delta k_{|i-j|}=|k_i-k_j|$. Red lines are the corresponding fits through Eq.~\eqref{eq:ProbTransv}.}
\end{figure*}

\begin{figure*}
	\includegraphics[width=1\linewidth]{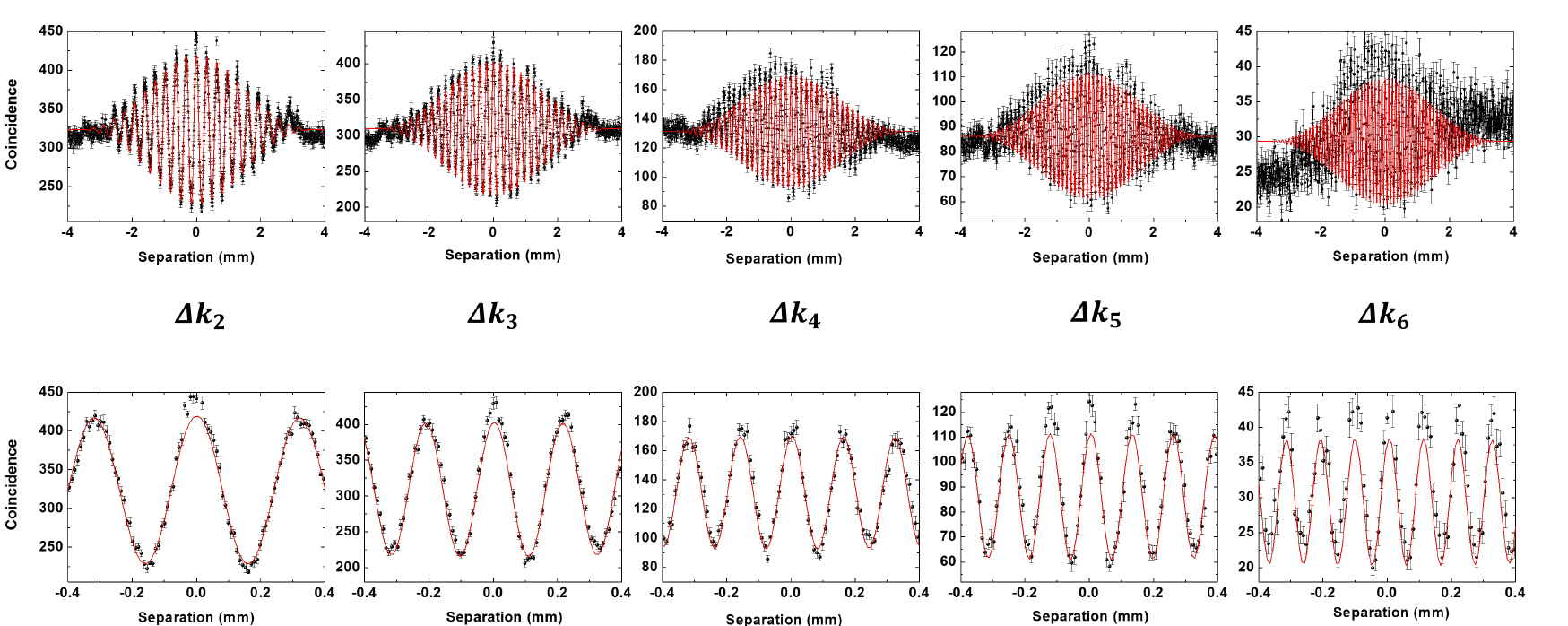}
	\caption{\label{MR AHOM} \textbf{Momentum resolved bunching.}
		Black points are the measured coincidences ($C^{b}_{i,j}$) for five different pixel pairs $i,j$ corresponding to the wavevector x-projections ($k_i,k_j$). The five plots show coincidence quantum beats $C^{b}_{i,j}$ between pixels corresponding to projections difference $\Delta k_{|i-j|}=|k_i-k_j|$.
		The lower panels show a magnification of the corresponding upper panels.
		The Red lines are the fits obtained through the Eq.~\eqref{eq:ProbTransv}.}
\end{figure*}

\section{Results}
\label{sec:results}
The measurement consists of acquiring the timestamps of the photons detected by each pixel of the SPAD array. 
By means of a script realized in the programming language Julia  we have computed the coincidences matrix $C^{B}_{i,j}$ between the $i$ and $j$ pixels around the same time difference (corresponding to the photons escaping the same BS output port, i.e.~photon bunching), and the coincidence matrix $C^{A}_{i,j}$ at 6 ns time difference, representing the antibunching (different output ports). 
At a given transverse separation $\Delta x$, each $C^{A/B}_{i,j}(\Delta x)$ value is obtained as the average of $l=1,2,...,n_r$ repeated measurements ${}^l C^{A/B}_{i,j}(\Delta x)$ each of which lasts $\sim$ 0.5 seconds, so that we can obtain the experimental uncertainty $\delta C^{A/B}_{i,j}(\Delta x)$ associated with $C^{A/B}_{i,j}(\Delta x)$ as the mean squared error of $^lC^{A/B}_{i,j}(\Delta x)$ divided by $\sqrt{n_r}$, where $n_r=10$.

If $k$ is the projection of the wave vector $\textbf{k}$ on $x$-direction ($k \overset{\text{def}}{=} k_x$), the center of each pixel $i$ of the SPAD array corresponds to a given $k$, centered at $k_i$ ($i=1,...,8$), with a range $\delta k$ due to the pixel dimension.
Figure~\ref{MR HOM} shows the measured coincidences $C^{A}_{i,j}$ between photons escaping different BS output ports (anti-bunching) for four different pixel pairs $i,j$.
Since each pixel $i$ corresponds to a given wave vector, $k_i$, the coincidences between pixels $i,j$ correspond to the difference $\Delta k_{|i-j|}=|k_i-k_j|$. 
In particular, Fig.~\ref{MR HOM} shows the coincidence data, 
for $\Delta k_0$, $\Delta k_1$, $\Delta k_2$, $\Delta k_3$ (black point), together with the corresponding regressions (red curves). %
Similarly, Fig.~\ref{MR AHOM} shows the measured coincidences $C^{B}_{i,j}$ between photons escaping the same BS output port (bunching), for five different pixel pairs $i,j$. 
In Fig.~\ref{MR AHOM} the coincidence counts for $\Delta k_0$ is missing because the detectors are not number-resolved. 
Also, the coincidence counts for $\Delta k_1$ is not shown because was too noisy due to cross talk of adjacent pixels.
As shown in Eq.~\eqref{eq:ProbTransv}, a suitable function to fit the data is
\begin{equation}\label{eq:ResFit}
	C^{A/B}_{i,j, \text{fit}}\left(\Delta x\right) = \mathcal{N} \left( 1\mp\mathcal{V} ~\sinc^2\left(\frac{\Delta x\delta}{2}\right)\cos(\Delta k\Delta x)\right)
	\, ,
\end{equation}
where the amplitude $\mathcal{N}$, the visibility 
$\mathcal{V}$ ($\sim 0.3$), the transverse-momentum sensitivity $\delta$ ($\sim 1.7 \, \text{mm}^{-1}$) as well as the beating frequency $\Delta k$ ($\sim 9.8 \, |i-j| \, \text{mm}^{-1}$) are parameters estimated by the regression.
The small discrepancy between the fitted value of $\delta\simeq1.7mm^{-1}$ and its theoretical value $\delta_{th}\simeq 2\pi w/\lambda f\simeq 2$mm$^{-1}$ can be caused by an active area slightly smaller than the nominal one due to the lower efficiency in the peripheral regions of the pixel caused by edge effects.
Regarding the value of the fitted beating frequency $\Delta k$, it depends on separation between pixels of interest. Considering two adjacent pixel ($|i-j|=1$), for example, such separation corresponds exactly to the pixel pitch; using the 
mapping function $ \Delta k = 2\pi \Delta y/\lambda f$ with $\Delta y= 250\mu m$, $f=300mm$ one finds $\Delta k=9.85 mm^{-1}$ that 
is compatible with the value obtained through the fit. All the other beating frequencies can be retrieved as multiples of this fundamental separation.
We observe that the oscillations are still visible when the transverse separation ($\Delta x$) between the beams is larger than its own transverse width ($\sigma_x$), i.e.~$\Delta x > \sigma_x \simeq 35\um{\mu m}$ (measured with the intense beam and a CCD).
This means that the experimental estimation of $\Delta x$ is possible even for non-overlapping beams, as predicted by our model. 
The increasing trend with the translator's position in Figures \ref{MR HOM} and \ref{MR AHOM} particularly visible for large $\Delta k$ is probably due to a slight asymmetry in the alignment of the optics, which causes a deformation of the photon's spatial profile as the translator's position changes. Clearly, such effects are more significant in the outer regions (large $\Delta k$) of the detector implying that the effect is larger in the plots corresponding to large $\Delta k$ .

To determine the experimental uncertainty $\delta \Delta x_\text{exp}$ we have adopted maximum-likelihood estimation.
We have used the fitted function $C^{A/B}_{i,j,\text{fit}}$ as probability mass function (function generating the data $C^{A/B}_{i,j}$).
We have computed the log-likelihood function
\begin{equation}\label{eq:LogLike}
	\mathcal{L}(\Delta x)=\sum_{\langle i,j \rangle ; \, X=A,B} C_{i,j}^X\ln C^X_{i,j,\mathrm{fit}}(\Delta x) 
\end{equation}
where the sum runs over all sampled pixel pairs $\langle i,j \rangle$ and over $X=A,B$ (about 64000 coincidence for each $\Delta x$)
and the uncertainty associated to the maximum-likelihood estimate $ \Delta x_\mathcal{L}$, after
\begin{equation}
	N=n_r \sum\limits_{\substack{\langle i,j \rangle ; \, X=A,B }}C_{i,j}^{X}   
\end{equation}
observed coincidences, through the error propagation formula
\begin{equation}\label{eq:UncRes}
	\delta \Delta x_{exp} = \sqrt{\sum_{\langle i,j\rangle ; \, X=A,B}
		\left(\frac{\dd \Delta x_\mathcal{L}}{\dd C_{i,j}^X} \, \delta C_{i,j}^X\right)^2} \, ,
\end{equation}
where $\delta C_{i,j}^{X}$  are the noise fluctuations that affect our experimental data $C_{i,j}^{X}$, and $\left|\frac{\dd \Delta x_\mathcal{L}}{\dd C_{i,j}^X}\right|$ are evaluated in appendix \ref{app:Likelihood}.

\begin{figure}
	\includegraphics[width=0.65\linewidth]{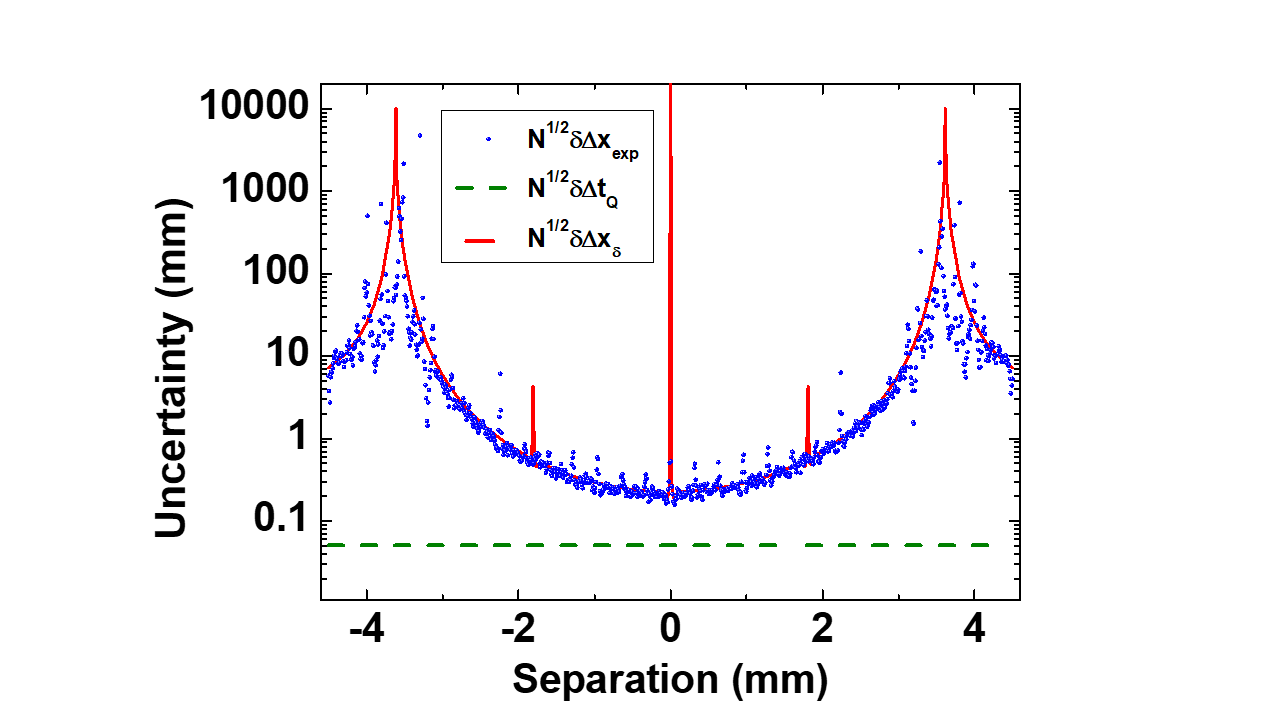}
	\caption{\label{Uncert} Uncertainties comparison: momentum resolved experimental uncertainty
		calculated through maximum-likelihood estimation (blue points), qCRB
		(green dashed line), FR CRB with finite momentum  resolution calculated in our model described in the supplemental material (red line). All three uncertainties are multiplied by $\sqrt{N}$. 
		The peaks of uncertainty visible in the plot of the CRB correspond to values of $\Delta x\simeq m \frac\pi\delta$, with integer $m=0,\pm1,\pm2$. Indeed, these are the stationary points of $\sinc^2(\frac{\Delta x\delta}{2})$.
	}
\end{figure}

Figure \ref{Uncert} shows the re-scaled error 
$\sqrt{N} \, \delta \Delta x_\text{exp}$ (blue points), the CRB for finite frequency resolution, computed through our model (red line), and qCRB (green dashed) per pair of observed photons.
In our experiment
we achieve high precision for 
$\Delta x  \gg \sigma_x\simeq 0.035$ mm, i.e.~in the absence of photon overlap at the beam splitter, as expected from our theoretical model.

\section{Conclusions}
\label{sec:conclusion}
We have demonstrated a measurement technique that exploits conjugate-variable passive resolution to achieve high-precision position estimation.
The technique relies on two-photon interferometry with independent photons and SPAD arrays.
Our approach stretches the operative range of Hong-Ou-Mandel interferometry beyond two-photon indistinguishability. 
In contrast to direct imaging, our scheme is not limited by the finite pixel pitch and, in principle, eliminates the need for magnifying objectives. This is a considerable advantage especially in imaging-related applications, 
such as biosensing and correlation plenoptic imaging~\cite{d2016correlation,di2020correlation,pepe2017diffraction}.
A clear example can be found in the last panel of Fig.~\ref{MR AHOM}: when the relative transverse displacement of one of the two optical paths is $ 20 \um{\mu m}$ the coincidence events count drops from 40 to 35.
Since this variation exceeds the size of our error bars, our system can effectively detect this small displacement, even though it is much smaller than the pixel pitch (250 $\mu m$).

Our results show that it is possible to observe quantum beats between independent photons whose wave packets are essentially non-overlapping.
The experiment presented here represents a proof-of-principle demonstration. 
We remark that there is ample room for technical improvements. For example, in our setup the acquisition quality already drops for $\Delta k_6$ 
and it limits the measurement of all $\Delta x$. In fact, the determination of the experimental $\Delta x$ is done using Maximum likelihood on all coincidences for all values of $\Delta k$ simultaneously. Thus, a low number of $\Delta k$ limits the achievable performance in terms of uncertainty.
This is not a fundamental limitation and can be overcome in future works.
Other technical improvements that will enhance the observed phenomenology include 
increasing the number of pixels and an boosting the performances of the SPAD arrays. Both these improvements are achievable with current or near future single-photon detection technologies.

\acknowledgements{
This work has received funding from the: 
European Union -- Next Generation EU:
NRRP Initiative, Mission 4, Component 2, Investment 1.3 -- D.D. MUR n. 341 del 15.03.2022 -- Next Generation EU (PE0000023 ``National Quantum Science and Technology Institute'');
European Union -- Next Generation EU, Missione 4 Componente 1, PRIN 2022, project title ``QUEXO'', CUP: D53D23002850006;
and the Italian Space Agency (Subdiffraction
Quantum Imaging ``SQI'' 2023-13-HH.0).
}

\appendix

\section{Evaluation of the quantum Fisher information}
\label{sec:appendix}

In this appendix we will evaluate the quantum Fisher information $H$ associated with the estimation of the transversal separation $\Delta x$ discussed in the main text.

Let us consider two coherent states $\ket{\alpha}_{j}$, $j=1,2$ of the type
\begin{equation}
	\ket{\alpha}_{j}=\e^{-\frac{\abs{\alpha}^2}{2}}\sum_{n=0}^\infty \frac{\alpha^n}{n!}\hat{a}_{j,x_j}^{\dag n}\ket{0},\qquad \hat{a}_{j,x_j}^{\dag}=\int\dd k\ \e^{-\ii k x_j} \psi(k)\hat{a}_j^{\dag}(k),
    \label{eq:CohState}
\end{equation}
where $\hat{a}^\dag_{j,x_j}$ is the photon operator that creates a photon at the $j$th input mode of the beam splitter in the transversal position $x_j$, $\hat{a}^\dag_j(k)$ creates a photon with transverse momentum $k$, and $\psi(k)$ is the transverse momentum wave-packet of a single photon.
In order to suppress single-photon interference effects and observe two-photon interference, we introduce a randomized phase $\phi_j$ that transforms the pure coherent state $\ket{\alpha}_{j}$ into the averaged state
\begin{multline}
	\hat{\rho}_{j}=\frac{1}{2\pi}\int_0^{2\pi}\dd \phi_j\ \ketbra{\e^{\ii\phi_j}\alpha}_{j}=\frac{1}{2\pi}\int_0^{2\pi}\dd \phi_j\ \e^{-\abs{\alpha}^2} \sum_{n,m=0}^\infty \frac{\alpha^{n}\alpha^{*m}}{n!m!}\e^{\ii\phi_j(n-m)}\hat{a}_{j,x_j}^{\dag n}\ketbra{0}{0}\hat{a}_{j,x_j}^{m}\\
    =\e^{-\abs{\alpha}^2}\sum_{n=0}^\infty \frac{\abs{\alpha}^{2n}}{n!^2}\hat{a}_{j,x_j}^{\dag n}\ketbra{0}{0}\hat{a}_{j,x_j}^{n}=\sum_{n=0}^\infty p_n \ketbra{n}{n}_{j,x_j},\quad p_n=\e^{-\abs{\alpha}^2} \frac{\abs{\alpha}^{2n}}{n!}
    \label{eq:randomization}
\end{multline} 
We notice that the state in Eq.~\eqref{eq:randomization} is a classical sum of states that remain orthogonal independently of the value of $x_j$, each with a different number of photons.
If we assume that only $2$-photon events are observed through post-selection from the global state $\hat{\rho}_1\otimes\hat{\rho}_2$, after normalization, the post-selected global state becomes
\begin{equation}
	\hat{\rho}^{2ph}=\frac{1}{2}\ketbra{1}_{1,x_1}\otimes\ketbra{1}_{2,x_2}+\frac{1}{4}\ketbra{2}_{1,x_1}\otimes\ketbra{0}_{2,x_2}+\frac{1}{4}\ketbra{0}_{1,x_1}\otimes\ketbra{2}_{2,x_2}
    \label{eq:2phState}
\end{equation}
The additivity over orthogonal sectors and over tensor products of the quantum Fisher information matrix~\cite{Liu2020} allows us to write the quantum Fisher information matrix $\mathcal{H}\equiv\mathcal{H}[\hat{\rho}^{2ph}]$ associated with $x_1,x_2$ as
\begin{equation}
	\mathcal{H}=\begin{pmatrix}
	    \frac{1}{2}H[\ket{1}_{1,x_1}] +\frac{1}{4}H[\ket{2}_{1,x_1}]& 0\\
        0 & \frac{1}{2}H[\ket{1}_{2,x_2}+\frac{1}{4}H[\ket{2}_{2,x_2}]]
	\end{pmatrix}
    \label{eq:QFIM2}
\end{equation}
where $H_{x_j}[\ket{n}_{j,x_j}]$ is the quantum Fisher information of the state with $n=1,2$ photons associated with $x_j$, $j=1,2$.
We can evaluate the terms inside Eq.~\eqref{eq:QFIM2} by employing the definition of the quantum Fisher information for a pure state
\begin{equation}
	H_{x_j}[\ket{n}_{j,x_j}]=4\left(\braket{\partial_j n}_{j,x_j}-\abs{\braket{n}{\partial_j n}_{j,x_j}}^2\right),\qquad \partial_j\equiv\frac{\dd}{\dd x_j}
    \label{eq:QFIDef}
\end{equation}
Given the definition of $\hat{a}^\dag_{j,x_j}$ in Eq.~\eqref{eq:CohState}, we can evaluate
\begin{align}
    \ket{n}_{j,x_j}&=\frac{\hat{a}_{j,x_j}^{\dag n}}{{\sqrt{n!}}}\ket{0}=\frac{1}{\sqrt{n!}}\int\dd k_1\dots\dd k_n\ \e^{-\ii(k_1+\dots+k_n)x_j}\psi(k_1)\dots\psi(k_n)\hat{a}^\dag_{j}(k_1)\dots\hat{a}^\dag_{j}(k_n)\ket{0}\\
	\ket{\partial_j n}_{j,x_j}&=\frac{1}{\sqrt{n!}}\int\dd k_1\dots\dd k_n\ -\ii(k_1+\dots+k_n)\e^{-\ii(k_1+\dots+k_n)x_j}\psi(k_1)\dots\psi(k_n)\hat{a}^\dag_{j}(k_1)\dots\hat{a}^\dag_{j}(k_n)\ket{0}
\end{align}
and thus, employing the commutation relation $[\hat{a}_j(k),\hat{a}^\dag_j(k')]=\delta(k-k')$ and the definition of quantum Fisher information in Eq.~\eqref{eq:QFIDef}, we get, for generic $n$
\begin{align}
	\braket{\partial_j n}_{j,x_j}&=\int\dd k_1\dots\dd k_n\ (k_1+\dots+k_n)^2\abs{\psi(k_1)\dots\psi(k_n)}^2=n\langle k^2\rangle+n(n-1)\langle k\rangle^2\\
    \abs{\braket{n}{\partial_j n}_{j,x_j}}^2&=\abs{\int\dd k_1\dots\dd k_n\ (k_1+\dots+k_n)\abs{\psi(k_1)\dots\psi(k_n)}^2}^2=n^2\langle k\rangle^2\\
    H_{x_j}[\ket{n}_{j,x_j}]&=4n(\langle k^2\rangle-\langle k\rangle^2)=4n\sigma_k^2,
    \label{eq:QFIN2}
\end{align}
where $\langle\cdot\rangle$ denotes the average over the single-photon transverse momentum distribution $\abs{\psi(k)}^2$.
We can use the expression of $H_{x_j}[\ket{n}_{j,x_j}]$ in Eq.~\eqref{eq:QFIN2} for $n=1,2$ to evaluate the terms in the quantum Fisher information matrix in Eq.~\eqref{eq:QFIM2}, ultimately yielding
\begin{equation}
	\mathcal{H}=4\sigma_k^2I_2,\qquad \mathcal{H}^{-1}=\sigma_x^2I_2
\end{equation}
with $I_2$ the $2\times2$ identity matrix, and $\sigma_x=1/2\sigma_k$ width of the spatial transverse wave-packet.
We can find the quantum Fisher information $H$ associated with $\Delta x$ by performing a change of parametrization $\Delta x=\Delta x(x_1,x_2)=x_1-x_2$ and introducing the Jacobian vector $J=(\partial_1 \Delta x, \partial_2 \Delta x)=(1,-1)$ of this reparameterization, yielding $H=J\mathcal{H}^{-1}J^T=2\sigma_x^2$. The quantum Cramér-Rao bound associated with $\Delta x$ after the observation of $N$ copies of the two-photon state in Eq.~\eqref{eq:2phState} thus reads
\begin{equation}
	\delta\Delta x_Q =\sqrt{\frac{J\mathcal{H}^{-1}J^T}{N}}=\sqrt{\frac{2}{N}}\sigma_x,
\end{equation}
as shown in the main text.

\section{Derivation of the momentum-resolved CRB for finite resolutions}
\label{app:derivation}

Given the probability function $P(x|\phi)$ describing the random variable $X$ taking values $x$ where $\phi$ is a parameter of the distribution, the Fisher information $F$ associated with the estimation of $\phi$ is by~\cite{Cramer1999}

\begin{equation}
	F = \mathbb{E}_X\left[\left(\frac{\partial}{\partial\phi}\ln P(X|\phi)\right)^2\right],
\end{equation}
where $\mathbb{E}_X$ is the expectation value over the random variable $X$.
Applied to our probability mass function $P_{A/B}^{k_{01},k_{02}}$ found in Eq. (4), we have
\begin{equation}
	F_{\delta}=\sum_{\substack{k_{01},k_{02}\\X=A,B}}\frac{1}{P_{A/B}^{k_{01},k_{02}}}\left(\frac{\partial P_{A/B}^{k_{01},k_{02}}}{\partial\Delta x}\right)^2,
	\label{eq:FIdeltaApp}
\end{equation}
where the summation over $k_{01},k_{02}$ is calculated over the central transverse momentum of all the pixels, which are enough  to cover the whole transverse-momentum distribution $f(k)$ of the photons.
According to Cramér-Rao bound the best precision achievable is thus
\begin{equation}
	\delta \Delta x_{\delta}=\frac{1}{\sqrt{N F_{\delta}}}
	\label{eq:crb}
\end{equation}
where $N$ is the number of repeated measurements.
The CRB plotted has been numerically retrieved by setting the central momenta equally distant $\{k_{01},k_{02}\}=\{n\delta,m\delta\}$ and summing over $n,m\in\{-50,-49,\dots,49,50\}$, which covers the whole spectrum $f$ for $\delta=1.7~\text{mm}^{-1}$ and $\sigma_x=\frac{1}{2\sigma_k}=0.035~\text{mm}$.

\section{Maximum-likelihood estimator and sensitivity $\delta \Delta x_{exp}$}
\label{app:Likelihood}

The maximum-likelihood estimator $\Delta x_\mathcal{L}(\{C_{i,j}^X\})$ is the value of $\Delta x$ that maximizes the likelihood function given the set of the observed coincidence  $\{C_{i,j}^X\}$
\begin{equation}
	\mathcal{L}(\Delta x)=\sum_{\substack{X=A,B\\{i,j}}}C_{i,j}^X\ln C^X_{{i,j},\mathrm{fit}}(\Delta x),
	\label{eq:DerLogApp}
\end{equation}
with $i,j$ running over the set of pixel pairs, where $C^X_{{i,j},\mathrm{fit}}$ is assumed as the exact probabilities that generated the coincidence
$\{C_{i,j}^X\}$.
Consequently:
\begin{equation}
	0=\frac{\partial \mathcal{L}(\Delta x)}{\partial \Delta x}\Bigg\vert_{\Delta x=\Delta x_\mathcal{L}}=\sum_{\substack{X=A,B\\{i,j}}}C_{i,j}^X\frac{\partial}{\partial\Delta x}\ln C^X_{{i,j},\mathrm{fit}}(\Delta x)\Bigg\vert_{\Delta x=\Delta x_\mathcal{L}}.
\end{equation}

Remembering that $\Delta x_\mathcal{L}\equiv\Delta x_\mathcal{L}(\{C_{i,j}^X\})$ depends on the outcomes and differentiating eq. S4 with respect to the outcomes $C_{i,j}^X$ we have
\begin{equation}
	0=\frac{\dd}{\dd C_{i,j}^X} \frac{\partial \mathcal{L}(\Delta x)}{\partial \Delta x}\Bigg\vert_{\Delta x=\Delta x_\mathcal{L}} =\frac{\partial^2 \mathcal{L}(\Delta x)}{\partial C_{i,j}^X\partial \Delta x}\Bigg\vert_{\Delta x=\Delta x_\mathcal{L}} +\frac{\dd \Delta x_\mathcal{L}}{\dd C_{i,j}^X}\frac{\partial^2\mathcal{L}(\Delta x)}{\partial \Delta x^2}\Bigg\vert_{\Delta x=\Delta t_\mathcal{L}}.
\end{equation}
then:
\begin{equation}
	\frac{\dd \Delta x_\mathcal{L}}{\dd C_{i,j}^X}=-\frac{\frac{\partial^2 \mathcal{L}(\Delta x)}{\partial C_{i,j}^X\partial \Delta x}}{\frac{\partial^2\mathcal{L}(\Delta x)}{\partial \Delta x^2}}\Bigg\vert_{\Delta x=\Delta x_\mathcal{L}}
\end{equation}
can be used to estimate the uncertainty $\delta\Delta x_{exp}$ associated with the maximum likelihood estimator through propagation of the uncertainties $\delta C_{i,j}^X$ of independent random variables $C_{i,j}^X$
\begin{equation}
	\delta \Delta x_{exp} = \sqrt{\sum_{\substack{X=A,B\\{i,j}}}\left(\frac{\dd \Delta x_\mathcal{L}}{\dd C_{i,j}^X}\delta C_{i,j}^X\right)^2}
\end{equation}

\bibliographystyle{unsrt}
\bibliography{MomentumResolvedHOM}

\end{document}